\documentclass[a4paper,twocolumn,showpacs,preprintnumbers,amsmath,amssymb]{revtex4}

\usepackage{graphicx,epsfig,subfigure}
\usepackage{dcolumn}
\usepackage{bm,amssymb,amsmath}

\usepackage{pstricks}

\newcommand{\reff}[1]{(\ref{#1})}
\newcommand{\re}{\text{e}}
\newcommand{\ri}{\text{i}}
\renewcommand{\vec}{\boldsymbol}

\newcommand{\R}{{\mathbb R}}
\newcommand{\C}{{\mathbb C}}

\def\ga{\gamma}\def\th{\theta}
\def\al{\alpha}\def\sig{\sigma}
\def\ds{\displaystyle}
\def\pa{{\partial}}
\def\eps{\epsilon}\def\epsi{\epsilon}
\def\ra{\rightarrow}

\newcommand{\barr}{\begin{array}}\newcommand{\earr}{\end{array}}
\newcommand{\bpm}{\begin{pmatrix}}\newcommand{\epm}{\end{pmatrix}}
\newcommand{\bsm}{\left(\begin{smallmatrix}}
\newcommand{\esm}{\end{smallmatrix}\right)}
\newcommand{\ba}{\begin{array}}\newcommand{\ea}{\end{array}}
\def\dd{\, {\rm d}}\def\ri{{\rm i}}

\def\Om{\Omega}
\newcommand{\bi}{\begin{itemize}}\newcommand{\ei}{\end{itemize}}
\newcommand{\ben}{\begin{enumerate}}\newcommand{\een}{\end{enumerate}}
\newcommand{\bce}{\begin{center}}\newcommand{\ece}{\end{center}}

\begin{document}

\title{Statistics for surface modes of nanoparticles with 
shape fluctuations}
\author{Felix R\"uting}
 \email{rueting@theorie.physik.uni-oldenburg.de}
\affiliation{%
Institut f\"ur Physik, Carl von Ossietzky Universit\"at, D-26111 Oldenburg
}

\author{Hannes Uecker}
\email{hannes.uecker@uni-oldenburg.de}
\affiliation{
Institut f\"ur Mathematik, Carl von Ossietzky Universit\"at, D-26111 Oldenburg
}


\begin{abstract}
We develop a numerical method for approximating the surface modes of sphere-like
nanoparticles in the quasi-static limit, based on an expansion of (the angular
part of) the potentials into spherical harmonics. Comparisons of the results
obtained in this manner with exact solutions and with a perturbation ansatz
prove that the scheme is accurate if the shape deviations from a sphere are not too
large.  The method allows fast calculations for large numbers of particles, and
thus to obtain statistics for nanoparticles with random shape fluctuations. As
an application we present some statistics for the distribution of resonances,
polariziabilities, and dipole axes for particles with random perturbations.
\end{abstract}

\maketitle

\section{Introduction}\label{sec:Intro}
The excitation of surface plasmons can cause strong interaction between light
and metallic nanoparticles. These plasmons are hybrid modes of the
electromagnetic field and the electron gas and are confined to the surface of
the particle. They give rise to an enhancement of the incident field by several
orders of magnitude~\cite{Kreibig_Vollmer,Bohren_Huffmann,Hao_Schatz}. This
enhancement enables a variety of applications ranging from the well-established
surface-enhanced Raman spectroscopy (SERS), which allows the detection of even a
single molecule~\cite{Nie_Emory,Kneipp}, to the emerging field of
plasmonics~\cite{Maier_Atwater,Ozbay}, which has led to prototypes of plasmonic
wave\-guides which effectuate optical energy transfer below the diffraction
limit~\cite{Maier_Atwater,Brongersma,Maier},

A simple realization of a plasmonic waveguide is a chain of metallic spheres. A
surface plasmon mode, typically a dipole mode, of the first sphere of the chain
is excited and the scattered field of this first particle excites a surface mode
in a sphere nearby and so the excitation can travel through the chain. There are
two crucial points for an efficient transport: The spatial structure of the
scattered field in the region of the neighboring sphere must allow for an
efficient excitation of the favored mode, and the overlap of the resonances of
the bordering spheres has to be big enough. Since any realization of a sphere
will deviate from an ideal one, thus introducing random fluctuations, it is
important to estimate the typical magnitude of such deviations which still allow
for an efficient transport. Therefore a simple and efficient numerical
method for approximating the surface modes of the sphere-likes
particles is needed.

There are many numerical methods for the determination of the
electromagnetic field in the present of nanosized scatterers, like the
finite difference time domain approach (FDTD), or so called 
{\em semi--analytical methods} based 
on expansions into special function systems like 
the multiple multipole method (MMP), 
or the discrete dipole approximation (DDA), see, e.g., 
\cite{Wriedt,Kahnert} for reviews. Essentially, 
all methods able 
to calculate the fields also allow to determine the surface modes. 
For example, in \cite{Noguez} the DDA is used for the
determination of the surface modes of nanoparticles, and in 
\cite{Abajo_Aizpurua,Abajo_Aizpurua2} a boundary
integral approach is proposed, which focuses on the surface modes, 
and has been used in \cite{Hohenester_Krenn} to determine the surface
modes of single and coupled spheres, cylinders and cube--like
nanoparticles. 

Here we use a semi-analytical approach based on an 
expansion of the potentials
into spherical harmonics, i.e., into modes $r^lY^m_l(\th,\phi)$ and
$r^{-(l+1)}Y^m_l(\th,\phi)$, and on the determination of the expansion
coefficients by the {\em physically motivated} projection of the boundary 
conditions onto the modes $r^lY^m_l(\th,\phi)$.  
See also \cite[Sec.~6]{mish02} for a review of various ways to determine
expansions from the boundary conditions in a variety of related
problems. For nonspherical particles, our approach corresponds to an
expansion into non--orthogonal modes and therefore is similar to the
usage of the Rayleigh hypothesis in the theory of scattering in
optics, where the scattered field at a perturbed interface is likewise
expanded in the solutions of the scattered field of the unperturbed
one~\cite{Nieto}. It is known that such expansion methods may fail if
the deviations from the ideal geometry become too large, see, e.g., 
\cite{Ramm} and the references therein. Nonetheless, 
additional to its simplicity and easy implementation 
the distinct advantage of our approach is its 
computational efficiency for nearly spherical particles. 
Thus it allows to calculate the
surface modes for many realizations of randomly distorted nanospheres
and so to statistically characterize their optical responses. 
 
The paper is organized as follows: The numerical method is explained in
Sec.~\ref{sec:scheme}, and validated in Sec.~\ref{sec:exact}, using the
cases of an ellipsoid and of a sphere with certain shape distortions as
benchmarks. In Sec.~\ref{sec:random} we give a statistical study
of the optical response of spheres and spheroids with random perturbations.

\section{The scheme}\label{sec:scheme}
We are interested in the surface modes of sphere-like nanoparticles,
described as some bounded domain $\Om\subset\R^3$, with boundary $\pa\Om$. We
restrict ourselves to particles that are small compared to the relevant
wavelengths and therefore employ the quasi-static approximation. In order to
determine their surface modes we consider an excitation from
infinity (described by the potential $\Phi_{ext}$), and calculate the potential
inside ($\Phi_-$) and outside ($\Phi_+$) the particle.
These potentials fulfill the Laplace equation
\begin{equation}
\label{eq:lap}
\Delta \Phi_{\pm}(\vec{x})=0 \quad \text{for } \vec{x}\not\in\pa\Om. 
\end{equation}
In addition, the boundary conditions on the surface $\pa\Om$ of the 
particle are
\begin{align}
\Phi_+(\vec{x})=\Phi_-(\vec{x}) \quad \text{for }\vec{x} \in \pa\Om, 
\label{eq:bc1}\\
\partial_n \Phi_+(\vec{x})=\epsilon \partial_n \Phi_-(\vec{x}) 
\quad \text{for }\vec{x} \in \pa\Om,\label{eq:bc2}
\end{align}
with the outward normal derivative $\partial_n$ and the permittivity $\epsilon$
of the particle.  The boundary condition \reff{eq:bc2} implies
that the particle is surrounded by vacuum, and is homogeneous, isotropic, and
non-magnetic; the dielectric properties are assumed to be local.

There are two different methods to determine the surface modes of a
nanoparticle from 
\reff{eq:lap}-\reff{eq:bc2}. The first is to assume that the potential
vanishes at infinity, i.e.\ to calculate the modes of the particle
that can be present without an external excitation. In this case the
problem can be reformulated as an eigenvalue problem with a real {\it
  plasmonic eigenvalue} $\epsilon$ for which a nontrivial solution of
Eqs.~\reff{eq:lap}-\reff{eq:bc2} with
$\lim\limits_{\|\vec{x}\|\ra\infty}|\Phi_+(\vec{x})|=0$
exists~\cite{Grieser}. In this interpretation the variable $\epsilon$
in \reff{eq:bc2} is not regarded as the generally complex
permittivity of the particle, but rather as a real eigenvalue. The
second method is to study the system \reff{eq:lap}-\reff{eq:bc2} with
an external excitation and thus to regard the $\epsi$ in 
\reff{eq:bc2} as the complex permittivity $\epsi(\omega)$ of the
particle. The system is then solved for different values of the
permittivity and a solution is called a surface mode if the field
inside and around the particle is enhanced. If the imaginary part of
the permittivity does not vary too much, then the enhanced fields
occur when the real part of $\epsi$ is equal to a plasmonic
eigenvalue. Thus the terms {\it plasmonic eigenvalue} and {\it
  resonant value} are closely related and will be used
interchangeably. In general there will be a difference in the number
of eigenvalues and resonant values. While there is a infinite number
of eigenvalues, the used excitation will choose some of these eigenvalues
and only for these an enhanced field will appear.
 
 We study the response of a particle to an external field and 
  assume that the potential at infinity equals the potential of the excitation;
\begin{equation}
\label{eq:bc3}
\lim_{\|\vec{x}\|\ra\infty}|\Phi_+(\vec{x})-\Phi_{ext}(\vec{x})|=0. 
\end{equation}
The basic idea is to expand the potential inside and outside the particle into
spherical modes which automatically fulfill the Laplace equation \reff{eq:lap},
\begin{equation}
\label{eq:exp_in}
\Phi_-(\vec{x})=\sum\limits_{l=0}^{\infty} \sum_{m=-l}^l\alpha_{l,m} r^l 
Y^m_l(\theta,\phi), 
\end{equation}
and $\Phi_+=\psi_++\Phi_{ext}$ with 
\begin{equation}
\label{eq:exp_out}
\psi_+(\vec{x})=
\sum\limits_{l=0}^{\infty} \sum_{m=-l}^l\beta_{l,m}r^{-(l+1)}Y^m_l(\theta,\phi)
\end{equation}
 and 
$\Phi_{ext}(\vec{x})=\sum\limits_{l=0}^{\infty} \sum\limits_{m=-l}^l
\gamma_{l,m} r^lY^m_l(\theta,\phi)$. 
Here $\vec{x}=r\big(\cos(\phi) \sin(\theta), \sin(\phi) \sin(\theta),
  \cos(\theta)\big)^t$ and the familiar spherical harmonics are denoted by
  $Y^m_l(\th,\phi)=P^m_l(\cos \theta) \re^{\ri m \phi}$: For $m{\ge}0$, the
  associated Legendre polynomials are $\ds P_l^m(s)=c_l^m\frac 1 {2^l
  l!}(1{-}s^2)^{m/2} \left(\frac{{\rm d}}{{\rm ds}}\right)^{l+m}(s^2{-}1)^l$
  with the scale factors $c^m_l=(-1)^m\sqrt{\frac{2l+1}{4 \pi}}
  \sqrt{\frac{(l-m)!}{(l+m)!}}$, and $P^m_l:=(-1)^m P^{-m}_l$ for $m<0$.

From \reff{eq:bc3} the
coefficients $\gamma_{l,m}$ are defined by the excitation
potential. Thus it remains to calculate the coefficients
$\alpha_{l,m}$ and $\beta_{l,m}$ from \reff{eq:bc1}
and~\reff{eq:bc2}.  In order to do this we use the following numerical
scheme. First we truncate to $|l|\le N$ such that 
$(N+1)^2$ coefficients $\alpha_{l,m}$
and $\beta_{l,m}$ have to be calculated. To get the required $2
(N+1)^2$ equations we project the boundary conditions~\reff{eq:bc1}
and~\reff{eq:bc2} onto the modes $r^lY^m_l(\theta,\phi)$ with degree
equal to or less than $N$. In detail, we require
\begin{align}
\label{m1}
&\int_{\pa\Om}(\Phi_--\psi_+)\,r^lY^m_l(\theta,\phi)\dd S\\
&=\int_{\pa\Om} \Phi_{Ext} \,r^lY^m_l(\theta,\phi)\dd S,\nonumber 
\\
\label{m2}&\int_{\pa\Om}(\eps\pa_n\Phi_--\pa_n\psi_+)\,r^lY^m_l(\theta,\phi)\dd S\\&=\int_{\pa\Om} (\pa_n\Phi_{Ext}) \,r^lY^m_l(\theta,\phi)\dd S.\nonumber
\end{align}
This yields a system of the form 
\begin{gather}
(M_1+\eps M_2)U=M_3G\label{lsys} 
\end{gather}
where $M_1,M_2, M_3\in\C^{2(N+1)^2\times 2(N+1)^2}$ are matrices 
which depend only on the geometry of the particle, $G\in \C^{2(N+1)^2}$ 
depends only on the $\ga_{lm}$, and $U\in \C^{2(N+1)^2}$ contains 
the unknown coefficients $\al_{lm}$ and $\beta_{lm}$. 

For a sphere $S_{r_0}$ of radius $r_0$ the
spherical harmonics $Y^m_l$ are an orthogonal (orthonormal if $r_0=1$) basis of
$L^2(\pa S_{r_0})$. Thus, \reff{lsys} decouples in the case of a sphere 
(becomes block
diagonal, see \ref{sec:app} for the precise structure) and yields
$(N+1)^2$ {\em exact} solutions of \reff{eq:lap}--\reff{eq:bc2}.  For a
perturbed sphere the $Y^m_l$ are no longer orthogonal, and the 
physically motivated idea of
projecting onto the modes $r^lY^m_l$ (instead of, e.g., projecting onto the
spherical harmonics $Y^m_l$, which at first might appear more natural) is as
follows: we may expect the fields to be localized near parts of $\pa\Om$ with
high curvature, and for perturbations of spheres (of radius $r_0$) these occur
most naturally for parts bulging {\em out}, i.e., for $r>r_0$. Thus, to minimize
the error, it appears reasonable to weight the spherical harmonics as test
functions in \reff{m1},\reff{m2} with $r^l$. This also complies with the
folklore rule to use the same functions as test functions and as ansatz
function. On the other hand, this rule is rather ambiguous here since we have
$(N+1)^2$ more ansatz functions, namely $r^{-(l+1)}Y^m_l$. However, these tend
to localize near ``flat'' parts of the perturbed sphere and are therefore less
useful as test functions.  We evaluated all three sets of test functions
$(r^lY^m_l)_{l,m}$, $(Y^m_l)_{l,m}$, and $(r^{-(l+1)}Y^m_l)_{l,m}$, against
available exact solutions for spheroids (see Sec. \ref{sec:exact}) and found
that $(r^lY^m_l)$ works best while $(Y^m_l)$ and even more so
$(r^{-(l+1)}Y^m_l)$ yield slower convergence.

As already pointed out in the Introduction, expansions like those given by
Eqs.~\reff{eq:exp_in} and \reff{eq:exp_out} are conceptually related to the
Rayleigh hypothesis, and may fail to converge if the deviation of the geometry
considered from the ideal geometry is too large. Therefore it is of great
importance to test the method against known exact solutions, and to control the
error. As shown below, for the present problem it turns out that already
moderate $N$ yield quite accurate results if the deviations from a sphere are
not too large. The achievable accuracy, however, also depends on the quantities
one wants to compute.  We find that typically $N=7$, which yields $M_j\in \C^{128\times 128}$,
is sufficient to calculate the resonant value of $\eps$ with high accuracy.

The generation of the matrices $M_j$ is the most expensive part of the scheme
since each entry requires the evaluation of surface integrals similar to the
ones in Eqs.~\reff{m1} and~\reff{m2}. However, once the matrices $M_j$ are 
generated,
for any given $\Phi_{ext}$ we only need to first calculate the coefficients
$\ga_{lm}$ and then solve some rather small linear system with given $\eps$. 
In
particular, the scheme allows for a fast parameter scan when solving the system
\reff{lsys} for different $G$, i.e., when rotating the incident field. 
For fixed $\eps$ we may also define a T--matrix 
$T=(M_1+\eps M_2)^{-1}M_3$ to obtain $U=TG$. 
The
calculation of the $\ga_{lm}$ is quite simple; for instance, for a constant
field in $(x,y,z)^t$-direction the coefficients are $\ga_{lm}=0$ for $l\neq 1$
and $\ga_{1,-1}=-x-\ri y$, $\ga_{1,0}=z$ and $\ga_{1,1}=x-\ri y$. 

We use the GNU Scientific Library~\cite{GSL} for the spherical harmonics and the
Cuba library~\cite{CUBA} for calculating the projection integrals
\reff{m1},\reff{m2}. The linear system \reff{lsys} is then solved with a
standard method from LAPack~\cite{LAPack}.

\section{Comparison with exact solutions and a 
perturbation ansatz}\label{sec:exact} As test cases for our scheme we consider
the surface modes of an ellipsoidal particle, for which an exact solution exists
\cite{Bohren_Huffmann}, and the case of a sphere with certain Gaussian
perturbations. For the latter we compare our results with the results of a recently developed
perturbation-theoretical ansatz~\cite{Perturbation}.

\subsection{Surface modes of an ellipsoid}\label{surf-sec} 
We start with a spheroid, i.e.~an ellipsoid with two identical semi--axes. The
spheroid is oriented such that the two identical axes are along the $x$- and
$y$- axis of the coordinate system. Furthermore, we choose the semi-diameter in
$x$- and $y$- direction as $1$\footnote{All lengths in the following will be dimensionless because there is no
natural length scale in the system and therefore the results are presented in a
scale invariant way.}. Thus the geometry of the test case is described by one
parameter $R$, the semi--axis in $z$-direction. As the dipole modes of a sphere
are excitable by a constant field, and we are interested primarily in
dipole-like modes, we use a constant incident field in the test cases.

Considering a constant incident field in $z$-direction, the exact
solution for the resonant value of the permittivity $\epsilon$ is~\cite{Bohren_Huffmann}
\begin{equation}
\epsilon(R)=1-\frac{1}{L(R)}
\end{equation}
with the depolarization coefficient $L(R)$ given by
\begin{eqnarray}
L(R)=\frac{R}{2}\int\limits_0^{\infty}\, 
\text{d}s\frac{1}{(s+R)^2\sqrt{2 (s+1)(s+R^2)}}.
\end{eqnarray}

Numerically we determine the resonances as follows: After calculating the
matrices $M_j$ we solve Eq.~\reff{lsys} for different values of
$\epsi$\footnote{In all calculations we use an imaginary part of the permittivity
of $10^{-2}$. This value is chosen because if we approximate the permittivity of
gold with a simple Drude model~\cite{Ashcroft}
$\epsilon(\omega)=1-\frac{\omega_p \tau}{\omega(\omega \tau + \text{i})}$ with
$\omega_p=1.4 \cdot 10^{16}\,\text{Hz}$ and $\tau=3 \cdot 10^{-14}\,\text{s}$
the permittivity for $\omega=8\cdot 10^{15}\,\text{s}^{-1}$ is about
$-2+0.01\text{i}$.}, and define the resonant value as that value of
$\epsilon$ which produces the biggest dipole-like near field, i.e., we
search the value of $\epsi$ which maximizes
$\sqrt{|\beta_{1,-1}|^2+|\beta_{1,0}|^2+|\beta_{1,1}|^2}$.

\begin{figure}[Hhbt]
\centering
  \epsfig{file=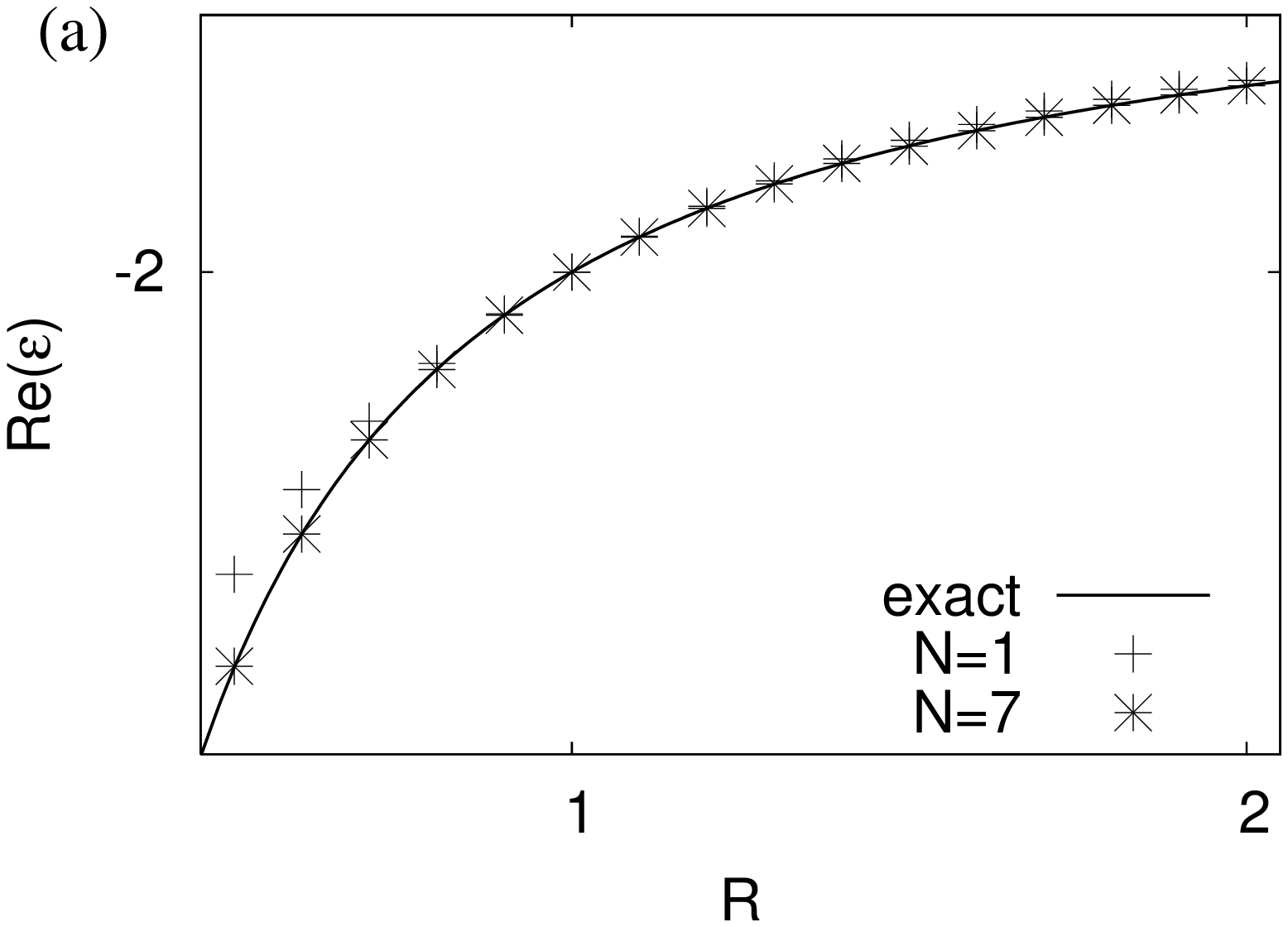, width=0.4\textwidth}
  \epsfig{file=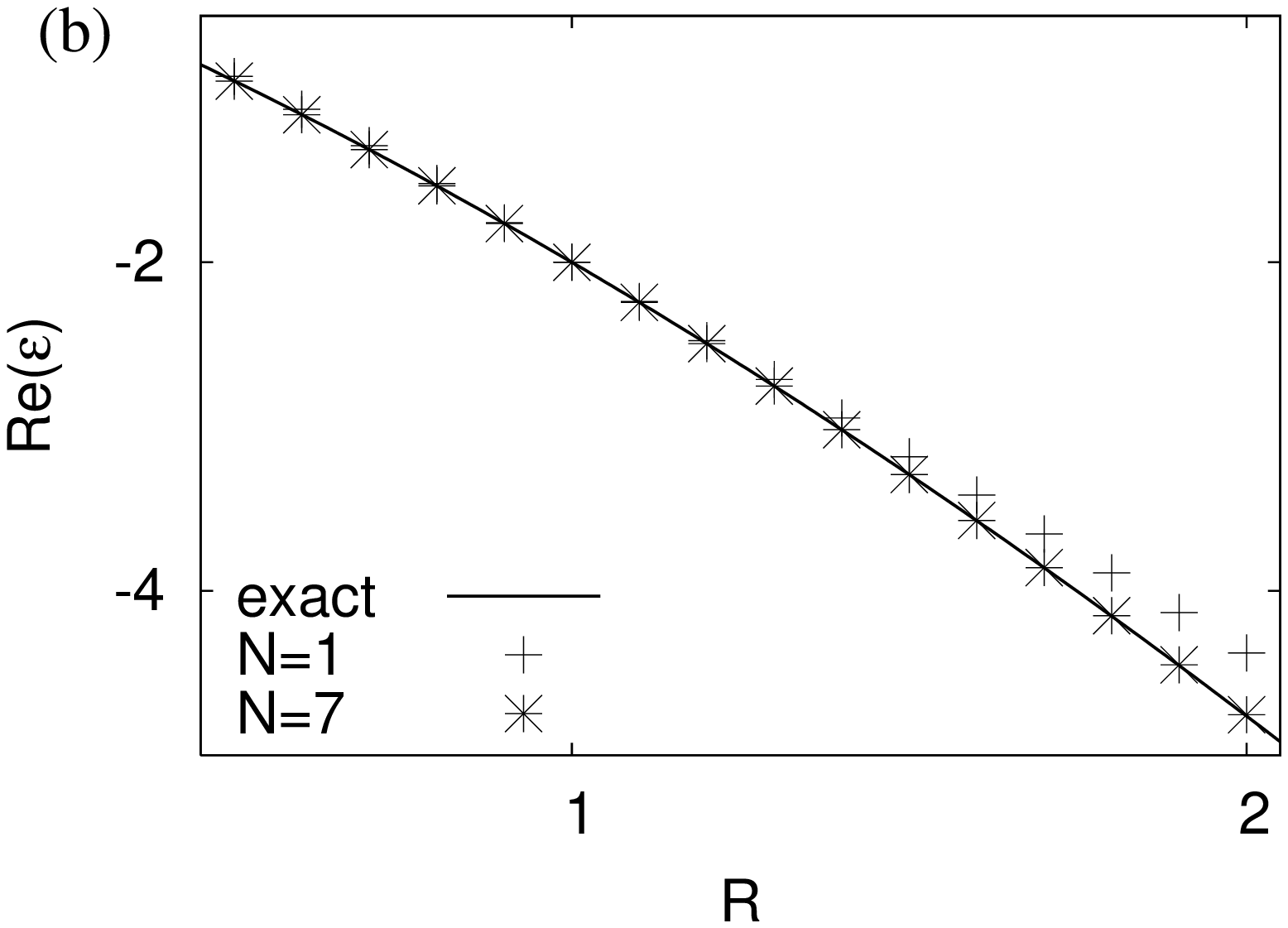, width=0.4\textwidth}
  \caption{\label{Fig:reso}Comparison of the exact resonant values for a
spheroid with the ones calculated within our approach for different $N$, and an
incident field oriented along the $x$-axis (a), and the $z$-axis (b). In the
later Fig.~\ref{fig:alpha1}, where we show the polarizability of a spheroids,
the dependence of the resonant values on $N$ can be seen in more detail.}
\end{figure}

In Fig.~\ref{Fig:reso} we compare the resonances thus obtained numerically with
the exact ones, for different $N$ and for two different incident fields. The
results are remarkably good, even for small $N$. Indeed, for an incident field
in $z$-direction and $R=1.5$, for example, the exact result figures as
$\epsilon\approx -3.29$, while the numerical procedure with $N=1$ gives
$\epsilon=-3.18$, and differs from the exact result by less than $10^{-2}$ with
$N=7$. Thus, for the calculation of the resonant values our method gives
accurate results even with only few spherical harmonics, and even when the shape
deviates significantly from that of a sphere.

\begin{figure}[Hhbt]
\centering
\epsfig{file=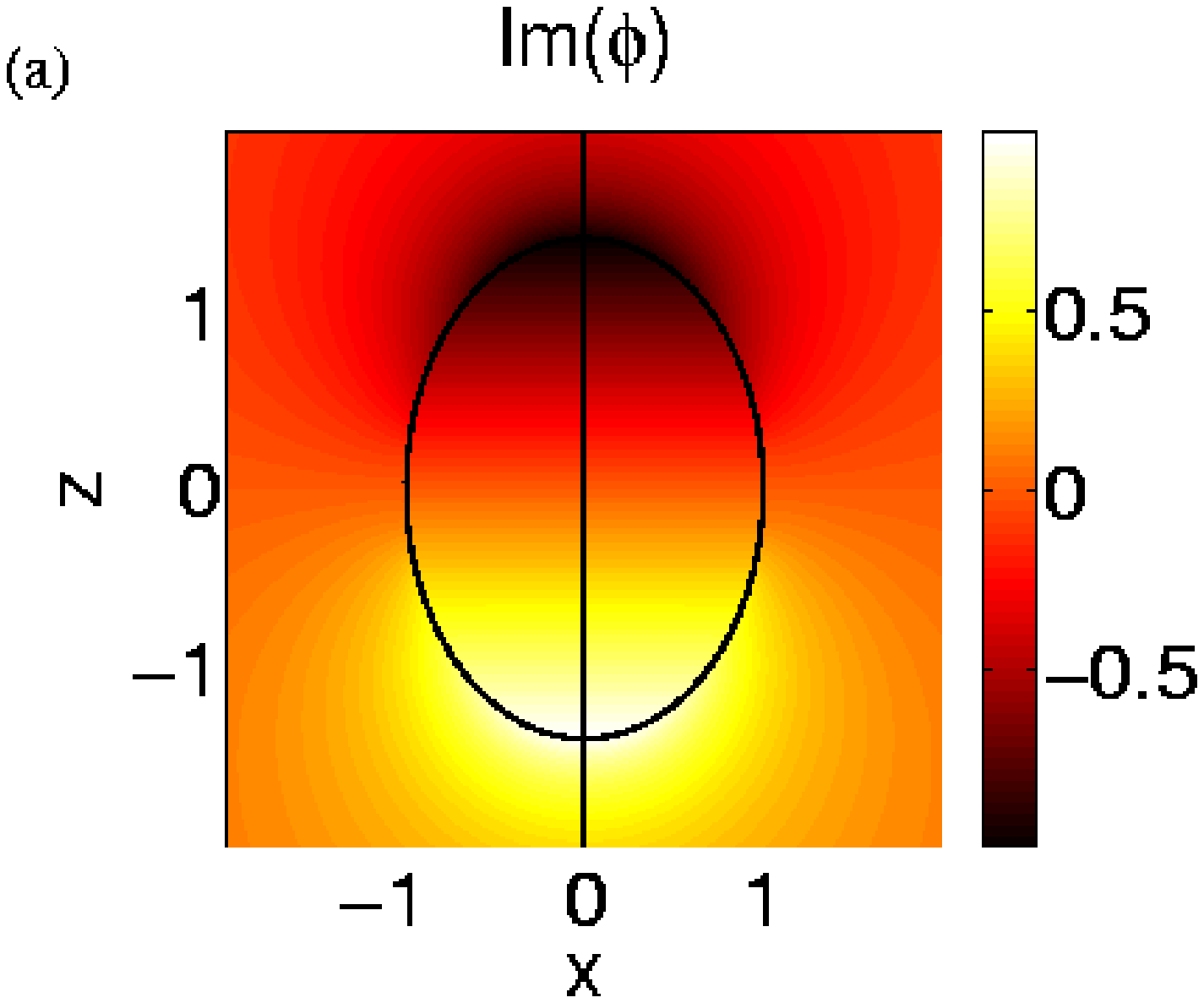, width=0.4\textwidth}
\epsfig{file=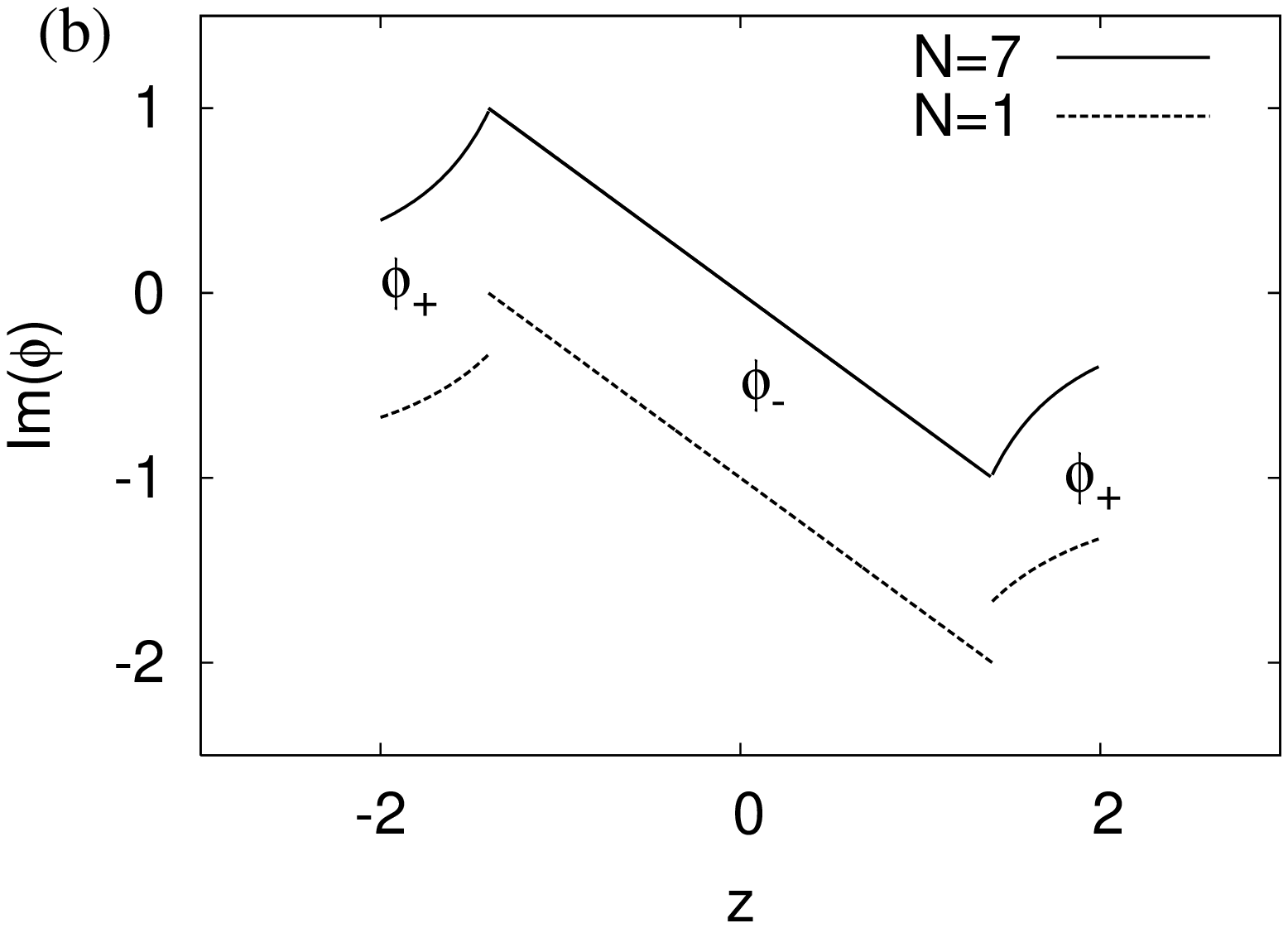, width=0.4\textwidth}
\caption{\label{fig:potential_ellip}(Color online) (a) Calculated potentials inside and around a
  spheroid for $R=1.4$ and $N=7$.  The imaginary part of the potential is
  plotted for the resonant value of the permittivity and the potential is
  normalized such that the maximum of the imaginary part is $1$. The shape of the
  spheroid is indicated by the black ellipse. The linescans in (b) are taken for
  $x=0$; the one with $N=1$ is shifted downwards by $1$ for clarity.}
\end{figure}

On the other hand, the method is in general not well suited for the calculation
of the fields with high accuracy if the deviations from the sphere become
large. For illustration, we first show in Fig.~\ref{fig:potential_ellip} (a) the
potentials inside and outside a spheroid with $R=1.4$, for an incident field in
$z$-direction. The linescan in (b) shows that along the particular line
indicated in (a) the boundary conditions \reff{eq:bc1}, \reff{eq:bc2} are
reasonably well fulfilled for $N=7$, while the jumps for $N=1$ indicate that in
this case $N=1$ is not sufficient, as expected.

\begin{figure}[Hhbt]
\centering
\epsfig{file=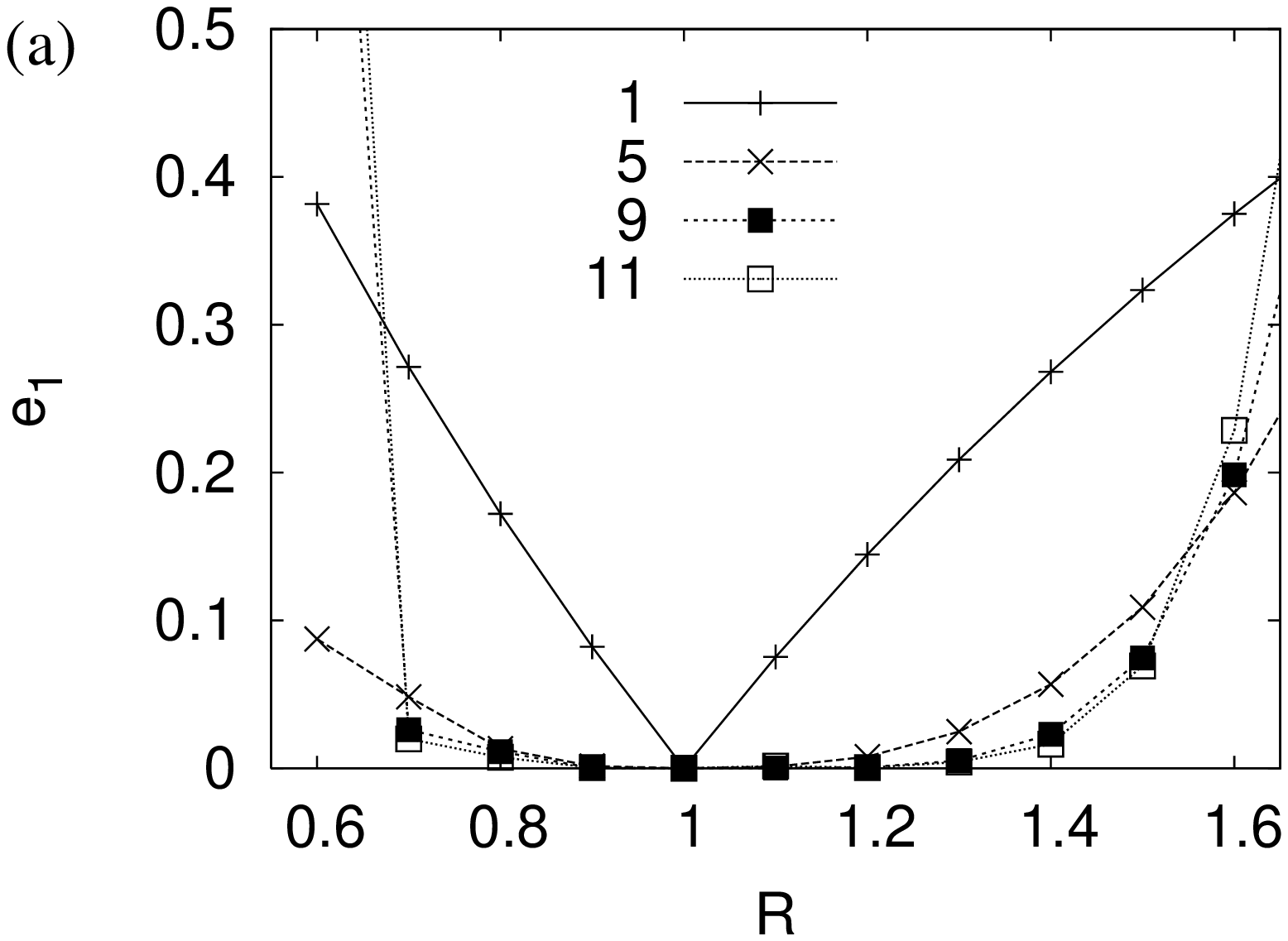, width=0.4\textwidth}
\epsfig{file=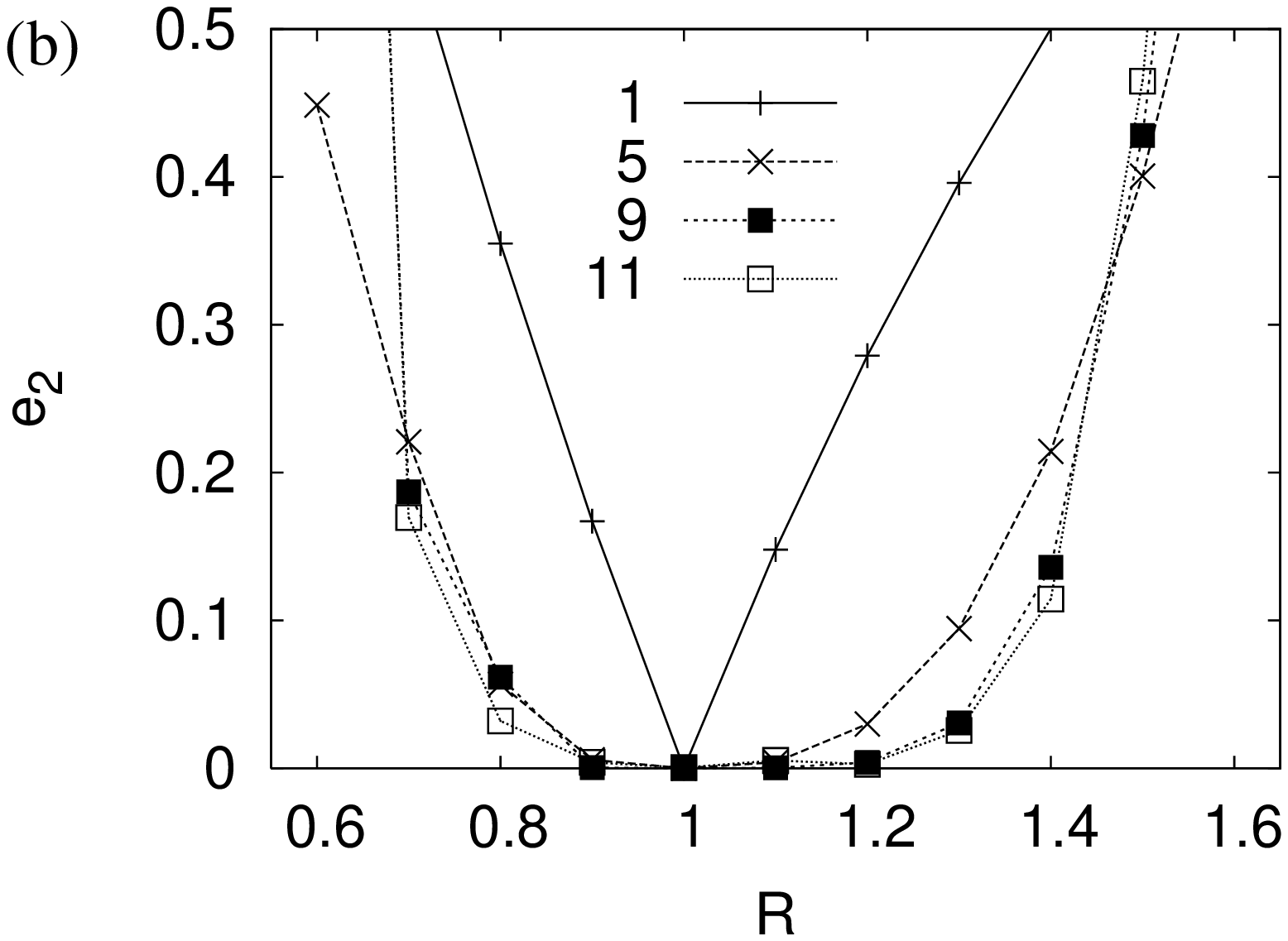, width=0.4\textwidth}
\caption{(a) $L_2$-norm of the boundary condition \reff{eq:bc1},
  and (b) of the condition \reff{eq:bc2}, for the indicated $N$. }\label{fig:rbcheck}
\end{figure}

From a systematic viewpoint, a more relevant basic check of the accuracy of the
potentials is the total relative error in Eqs.~\reff{eq:bc1},
\reff{eq:bc2}. In Fig.~\ref{fig:rbcheck} we plot the $L^2$--norms of these
errors, again as functions of $R$ and $N$, henceforth denoted by
$$
e_1:=\frac{2\|\Phi_+-\Phi_-\|}{\|\Phi_+\|+\|\Phi_-\|},\quad 
e_2:=\frac{2\|\pa_n\Phi_+-\eps\pa_n\Phi_-\|}
{\|\pa_n\Phi_+\|+\|\pa_n\Phi_-\|},
$$ where $\|f\|=(\int_{\pa\Om} |f|^2\dd S)^{1/2}$. For $R$ close to $1$, say
$0.8\le R\le 1.3$, we find that $e_1$ and $e_2$ are small and decrease
monotonously in $N$.  However, $e_{1,2}$ become large rather quickly when $R$
falls outside this range.  Moreover, they then no longer decay in $N$, which we
attribute to a failure analogous to that of the Rayleigh hypothesis for such
strongly deformed spheres.

Similar effects can be observed for various other test particles: The
approximation of resonant values of $\epsi$ is typically much better than
$e_{1,2}$. Thus the performance of the method strongly depends on what one wants
to compute.  As a rule of thumb we find that for $e_{1,2}\le 0.2$ we obtain very
good approximations of the resonant values, and also of the polarizabilities
and of the dipole axes, as shown below; these three observables are the
quantities we are mainly interested in. Therefore we have made sure that in all
calculations presented below we have $e_{1,2}< 0.1$.  For the solution shown in
Fig.~\ref{fig:potential_ellip} with $N=7$ we have $e_2$ slightly larger than
$0.1$, so that this solution would be discarded.

Another quantity of immediate physical interest is the polarizability $\alpha$
which relates the incident field ($\vec{E}_{ext}$) to the excited dipole moment
($\vec{p}_{ex}$). In general this quantity is a tensor, but for the case of a
spheroid with an incident field oriented along one of the principal axes the
polarizability is a scalar~\footnote{ Our method is also able to calculate the
full polarizability tensor by solving Eq.~\reff{lsys} for three independent incident
fields.}, i.e.,~$\vec{p}_{ex}=\epsilon_0 \alpha
\vec{E}_{ext}$.
\begin{figure}[Hhbt]
\centering
\epsfig{file=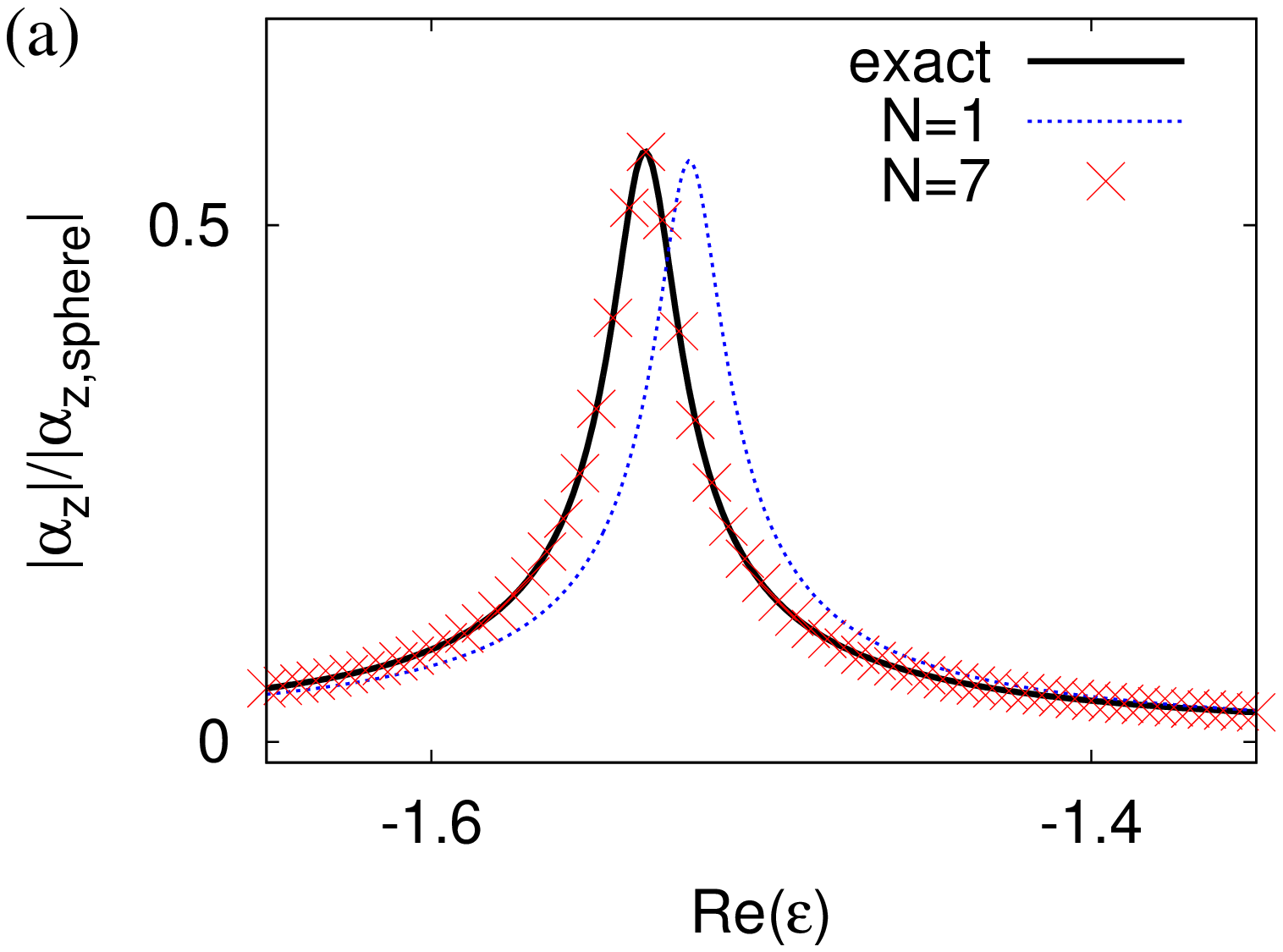, width=0.4\textwidth}
\epsfig{file=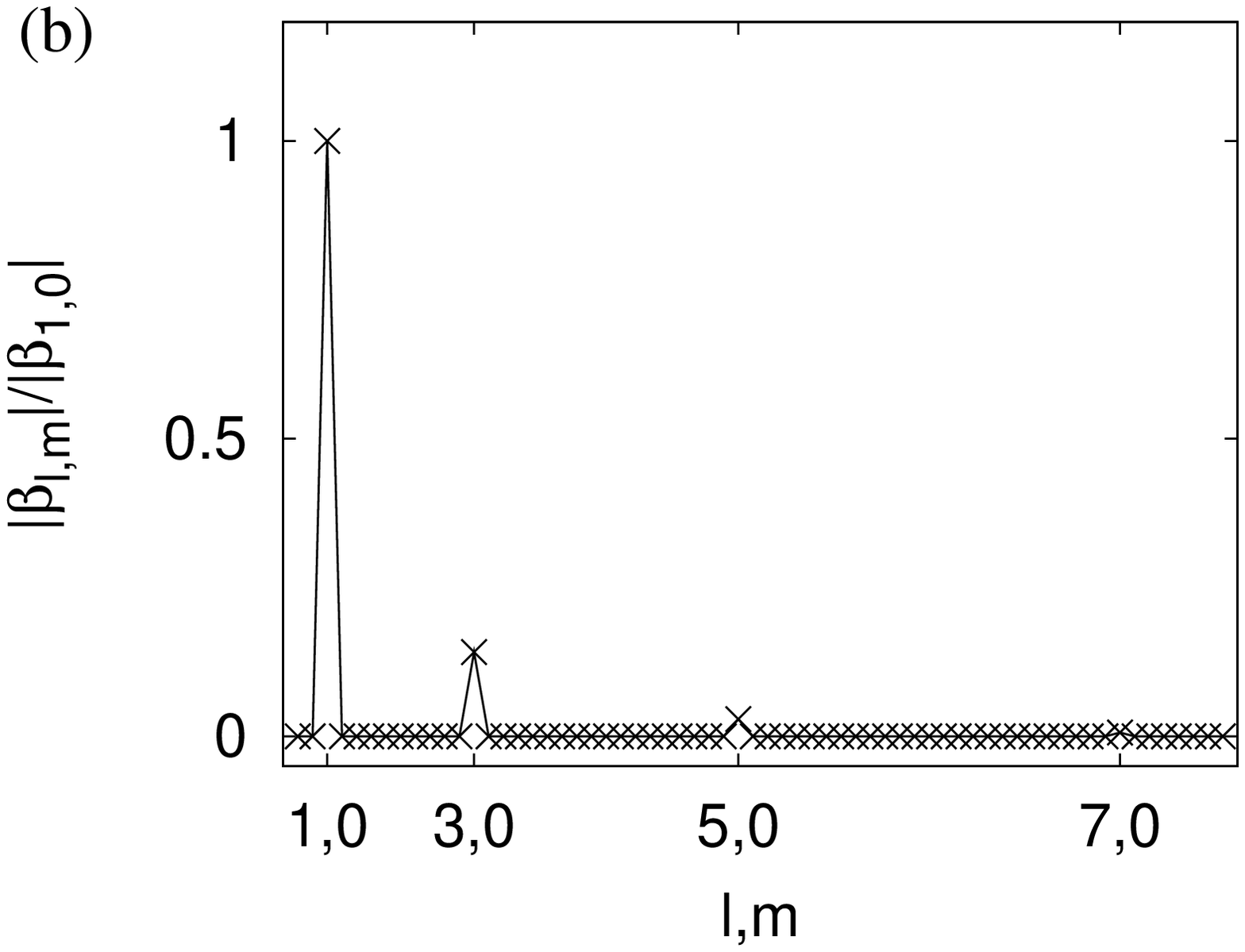, width=0.4\textwidth}
\caption{\label{fig:alpha1} (Color online) Comparison of the exact and
  numerically calculated polarizability of a spheroid with $R=0.8$
  (a). The polarizability is normalized to the one of a sphere with
  radius $1$ and a dielectric function of $-2+0.01\text{i}$. In the
  right panel (b) the coefficients of the spherical harmonics are
  presented for $N=7$ and a value of $\epsi$ close to the resonant
  value.}
\end{figure}
In Fig.~\ref{fig:alpha1}(a) the exact and the numerically calculated
polarizability of a spheroid with $R=0.8$ is depicted, more precisely
its absolute value for an incident field in $z$-direction,
i.e. $|\alpha_z|\propto\sqrt{|\beta_{1,-1}|^2+|\beta_{1,0}|^2+|\beta_{1,1}|^2}$.
The width and the magnitude of the polarizability are excellently
reproduced by our scheme and the agreement between the exact solution
and the numerical one is very good already for $N\geq 3$, and even for
$N=1$ the approximation already reproduces the shape of the resonance
quite well. Fig.~\ref{fig:alpha1}(b), which depicts the coefficients
$\beta_{l,m}$ for the expansion \reff{eq:exp_out}, shows that the
dominant part of the field stems from $l=1$, but higher spherical
harmonics also give non-negligible contributions.

We have also checked our scheme against the exact solution for an
ellipsoid with three different semi-diameters, and an incident field
that is not oriented along one of the principal axes. Again the
agreement between the exact solution \cite{Bohren_Huffmann} and our
results is very good as long as the semi-diameters are not too
different.

\subsection{Surface modes of a sphere with a perturbation}
In Sec.~\ref{sec:random} we perform a statistical analysis for the surface
modes of nanoparticles with a random shape, described by
\begin{equation}
\label{eq:gaussian}
r(\theta,\phi)=1+s \sum\limits_{i=1}^n h_i
\exp\left(-0.5\left(
\frac{{\rm dist}(\theta_i,\phi_i;\theta,\phi)}{w_i} \right)^2\right)
\end{equation}
where $s$ is a scaling factor, dist$(\theta_i,\phi_i;\theta,\phi)$ is
the Euclidean distance between two points on the unit sphere, one
specified by $\theta_i$ and $\phi_i$ and the other by $\theta$ and
$\phi$. In these later studies, $\theta_i$, $\phi_i$, $h_i$ and $w_i$
will be randomly distributed. Thus, the nanoparticles then are spheres
with $n$ Gaussian perturbations with height $h_i$ and width $w_i$. In
order to assess our method for such cases, we first
consider a particle with three perturbations, study how the
resonance is shifted by varying the scaling parameter~$s$, and compare
the results with those of a perturbation ansatz for the surface modes
of a nanoparticle~\cite{Perturbation}.

For the perturbation ansatz a geometry close to the given one is needed, for
which an exact solution for the surface modes exists. In the following that
geometry is called the ideal geometry, and the geometry for which the surface
modes are to be calculated is called the perturbed geometry. 
It is assumed that
the perturbation strength is described by a scalar parameter. In our case the ideal
geometry is the unit sphere and the parameter that characterizes the deviation
from the sphere is the scaling amplitude $s$. When making the perturbation ansatz the
resonant values $\epsilon(s)$ are expanded in a series with respect to $s$:
\begin{equation}
\epsilon(s)=\epsilon(0)+s\dot{\epsilon}(0)+\frac{s^2}{2}\ddot{\epsilon}(0)
+\cdots.
\end{equation}
In Ref.~\cite{Perturbation} an explicit formula for $\dot{\epsilon}$
is presented against which we can compare our numerical
results. Since the dipole mode of a sphere is threefold degenerate,
there are in general three different values of $\dot{\epsilon}$,
characterizing the three dipole-like modes.

\begin{figure}[Hhbt]
\centering
\epsfig{file=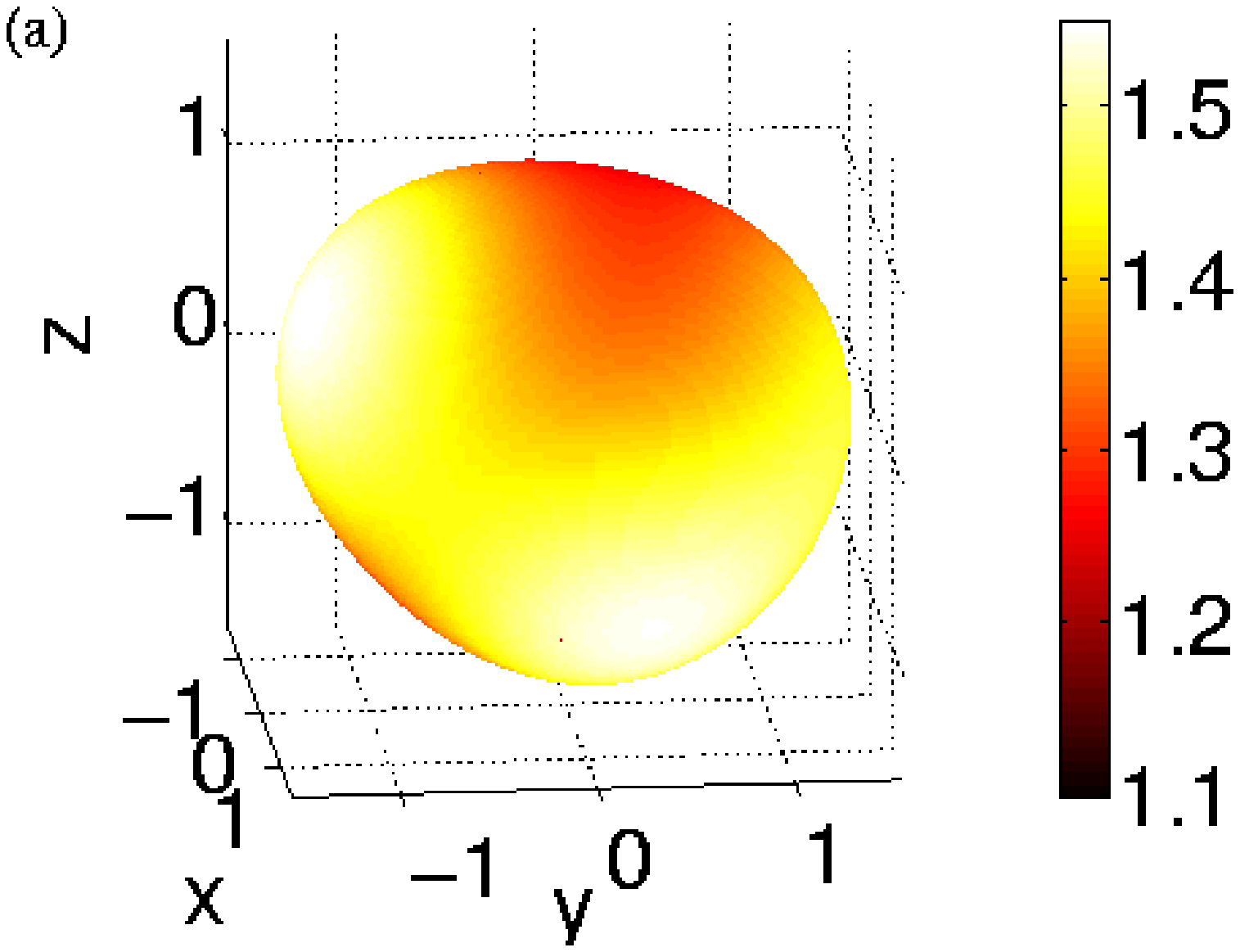, width=0.4\textwidth}
\epsfig{file=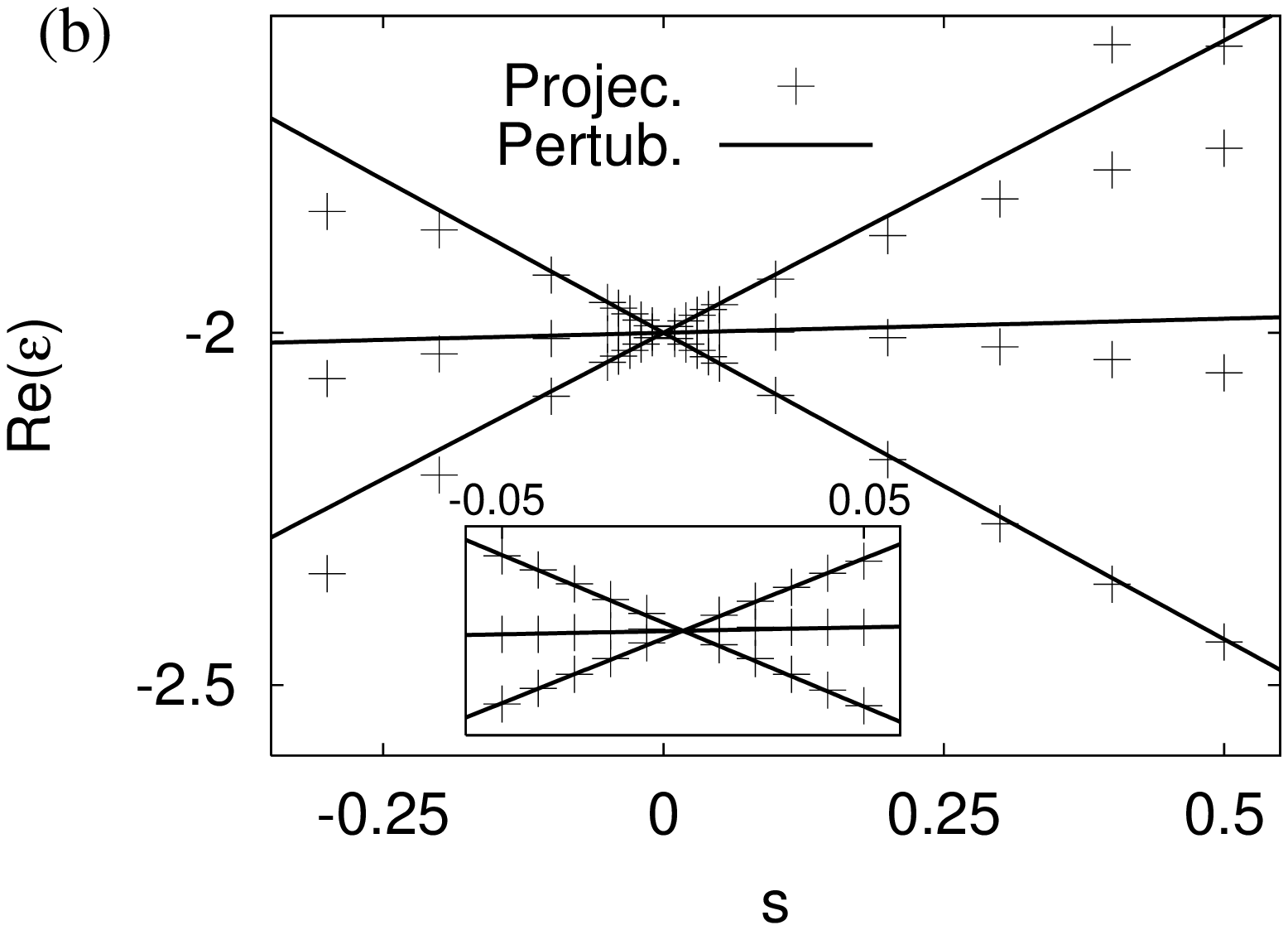, width=0.4\textwidth}
\caption{\label{fig:geometry3}(Color online) Sketch of the used perturbed 
sphere (a), and the corresponding resonant values of
  $\epsi$ (b). The plot (a) refers to a scaling factor of $s=0.5$; the color
  scale depicts the distance from a point on the surface to the center of the
  underlying unit sphere. The resonant values in (b) are calculated with $N=7$;
  the inset shows a blow-up near $s=0$.}
\end{figure}

The test particle with three Gaussian perturbations
is depicted in 
Fig.~\ref{fig:geometry3}(a), whereas (b) 
compares the results provided by \reff{lsys} 
with the perturbation method. We
only present the results for $N=7$ but remark that the results are stable for
$N\geq5$. There is a very good agreement between the projection and
the perturbation method for small $s$. As expected there are deviations for
larger scaling factors, because these are beyond the scope of the (first order)
perturbation method.

For $s\geq0.4$ more than three resonances occur, notably one with
a real part of the permittivity of about $-1.6$. By plotting the potential (or
alternatively inspecting the expansion coefficients $\beta_{l,m}$) we find that
these modes are shifted and perturbed quadrupole modes of the unperturbed
sphere, which due to the perturbations can be excited by a constant field.  For
the unperturbed sphere these occur at $\text{Re}(\epsilon)=-1.5$. Of
course these modes are in principle present in all cases, but it
depends on the respective geometry if these modes could be excited
effectively by a constant field.

To conclude the test of our scheme: Already with quite few spherical harmonics
(typically $N\le 7$), and even if the errors $e_{1,2}$ in the boundary
conditions are significant, the numerical data for the resonant values and the
polariziabilities provided by our method are stable and in remarkable agreement
with exact solutions (if available), or with the results of the perturbation
ansatz.

\section{Particles with randomly distributed Gaussian 
perturbations}\label{sec:random} We now assume that we are given a set
of nominally identical nanospheres which suffer from uncontrolled
shape imperfections induced during the fabrication process. The task
then is to characterize the optical properties of this set in a
statistical sense. Ideally, one would have at least a good idea how
the imperfections are distributed, based on an inspection of a
representative number of specimen, then generate a corresponding
ensemble, and compute the resonances of each individual member. Since
we are not considering any particular case, here we simply assume that
the random shape fluctuations correspond to Gaussian distortions as
described by Eq.~\reff{eq:gaussian}. We start with $n=4$, set $s=1$,
and choose randomly distributed $\theta_i$, $\phi_i$, $h_i$, and $w_i$
with $i=1,\ldots,4$.  The positions of the distortions, described by
the angles $\theta_i$ and $\phi_i$, are uniformly distributed on the
sphere, $h_i$ is normally distributed with a mean of $0.2$ and a
standard deviation of $0.1$; the normally distributed widths $w_i$
have a mean value of $\mu_w=0.7$ and a standard deviation of
$\sig_w=0.3$. Clearly, perturbations with a negative width $w_i$ are
discarded, and we admit only perturbations with $\frac{h_i}{w_i}\leq
2$ in order to avoid sharp peaks. Therefore each realization is a
sphere with four or less perturbations. 

In order to get significant results we generate 1000 realizations of the
perturbed sphere, calculate for each realization the matrices $M_i$, and solve
the system of equations \reff{lsys} for 100 different incident fields, i.e.~we
consider 100000 cases.  We choose $N=7$, and require $e_{1,2}\le 0.1$ as explained in
Sec.~\ref{surf-sec}, which has resulted in an ensemble with 42829 members.

\begin{figure}[Hhbt]
\centering
\epsfig{file=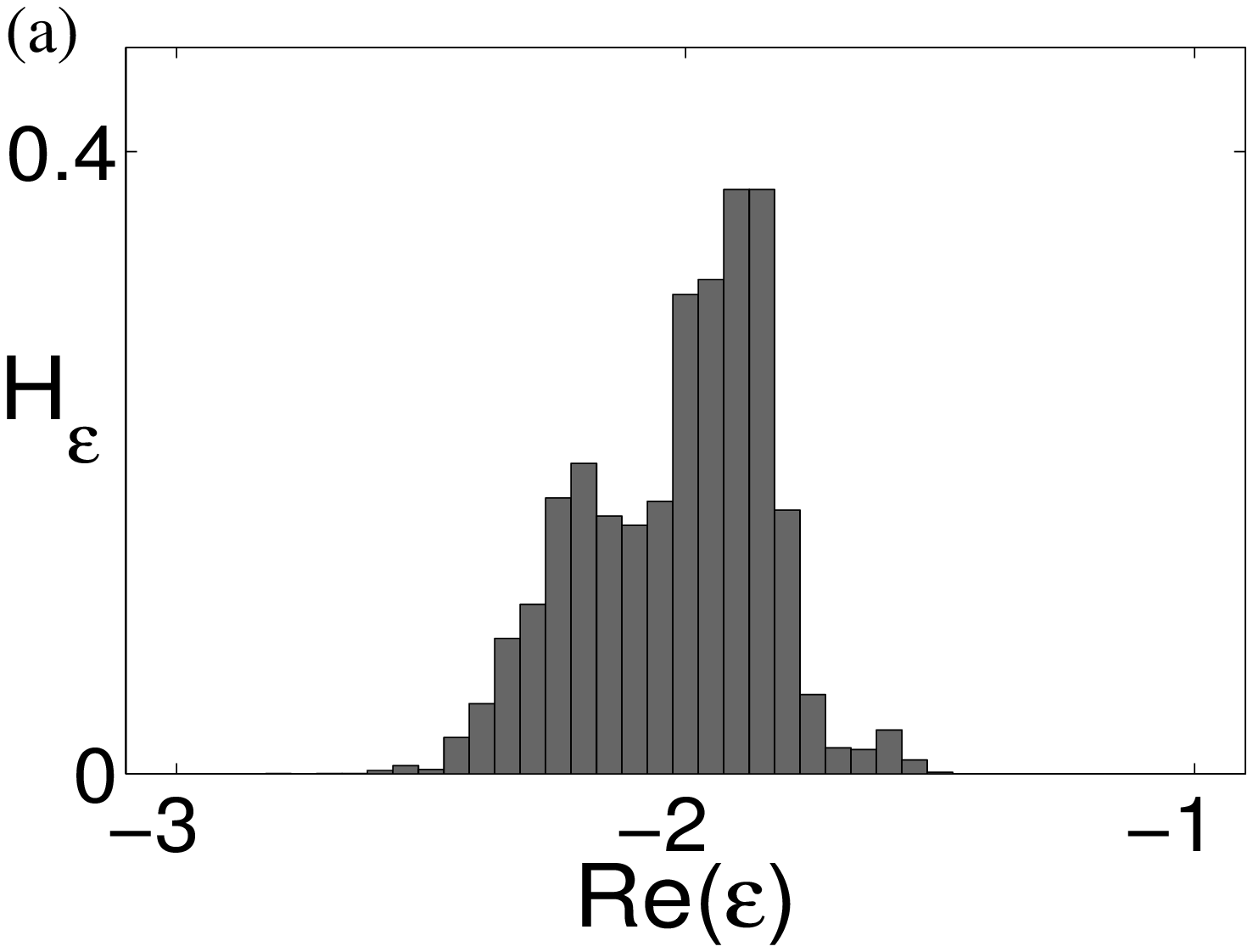, width=0.4\textwidth}
\epsfig{file=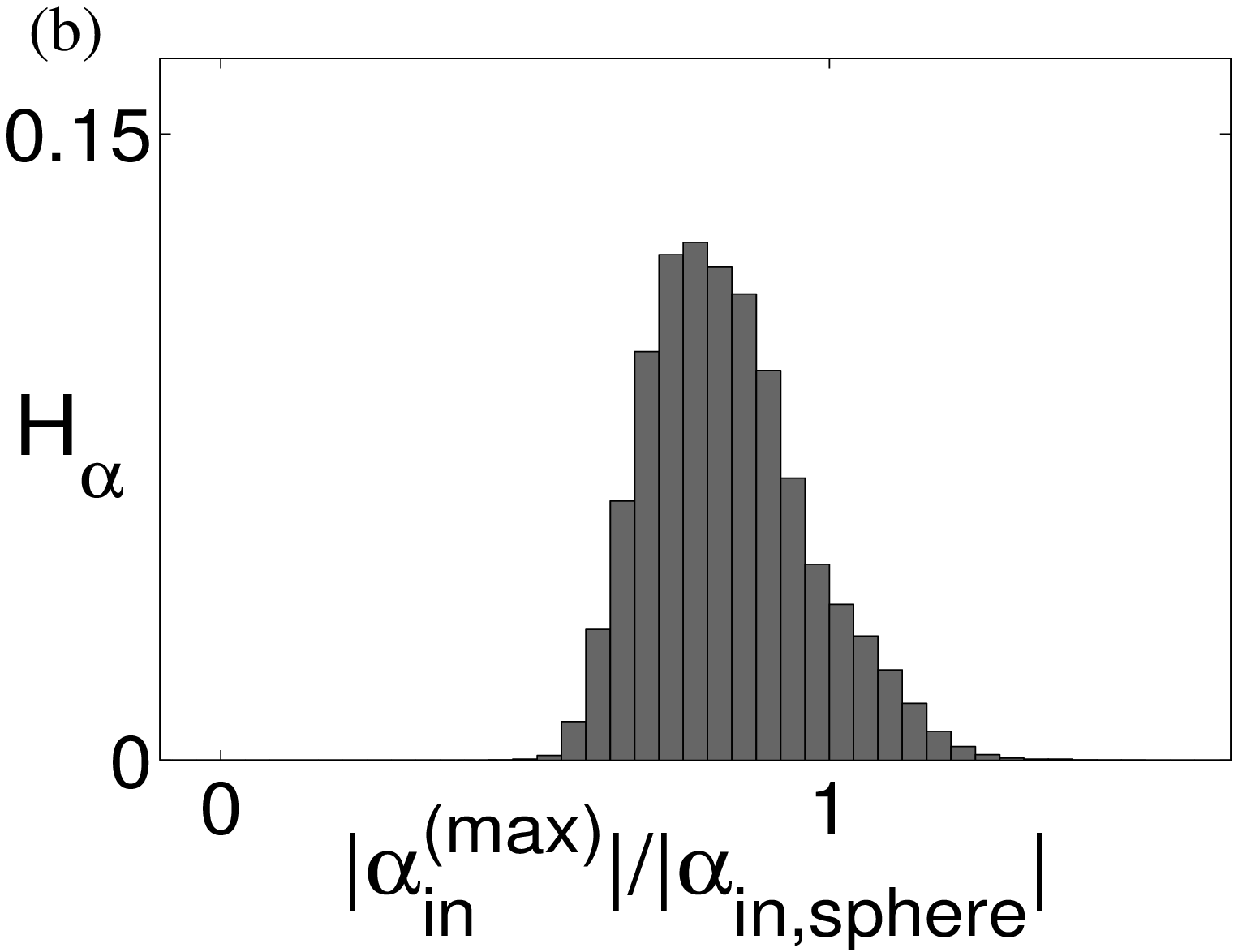, width=0.4\textwidth} 
\caption{Histogram of the resonant values Re$(\epsilon)$ 
  for the ensemble of perturbed spheres (a), and histogram of the magnitude of the 
maximum dipole
moment $\alpha_{in}^{(max)}$ (b).\label{fig:histo} }
\end{figure}

In Fig.~\ref{fig:histo}(a) we present a histogram of the resonant values of the
real part of the permittivity. Here we count a peak in the absolute value
$|\alpha_{in}|$ of the polarizability of a particle as a resonance if
$|\alpha_{in}|$ is at least $0.1$ times the value $|\alpha_{in,sphere}|$ of a
perfect sphere. $H_{\epsilon}$ denotes the number of the resonances in the
corresponding interval of $\text{Re}(\epsilon)$ divided by the number of
particles. In the following pictures $H_{\alpha}$ and $H_{\theta}$ are defined
in an analogous way. As seen in Fig.~\ref{fig:histo}(a) the shape fluctuations
result in a relatively broad distribution of the resonant values around the
unperturbed value $\text{Re}(\epsilon)=-2$, and the maximum is shifted to a
slightly bigger value.

Not only the position of the resonance is of interest, but also the magnitude of
the induced dipole moment. Therefore we present in Fig.~\ref{fig:histo}(b) a
histogram of this magnitude, considering for each ensemble member only the
resonance with the largest dipole moment. Due to the fact that the
polarizability of a sphere scales with its volume and a typical realization of
our perturbed spheres is somewhat bigger than a sphere with radius $1$, we
normalize the polarizability to that of a sphere with the same volume as the
perturbed one.  In the large majority of cases the polarizability of the
perturbed sphere is smaller than that of the unperturbed one. This is due to the
fact that for the sphere there are no principal axes. Therefore the
polarizability is independent of the direction of the incident field, and for
any field the induced dipole moment points into the direction of the incident
field. But if the symmetry of the sphere is destroyed by the perturbations,
there are three distinguished principal axes with different resonant values of
the permittivity. So for a fixed permittivity in general only one resonant
condition is matched and therefore essentially only the part of the incident
field that points into the direction of the corresponding principal axis
contributes significantly to the induced dipole moment. This results in an
induced dipole moment which typically is smaller than that of a perfect sphere,
and is not parallel to the incident field.

\begin{figure}[Hhbt]
\centering
\epsfig{file=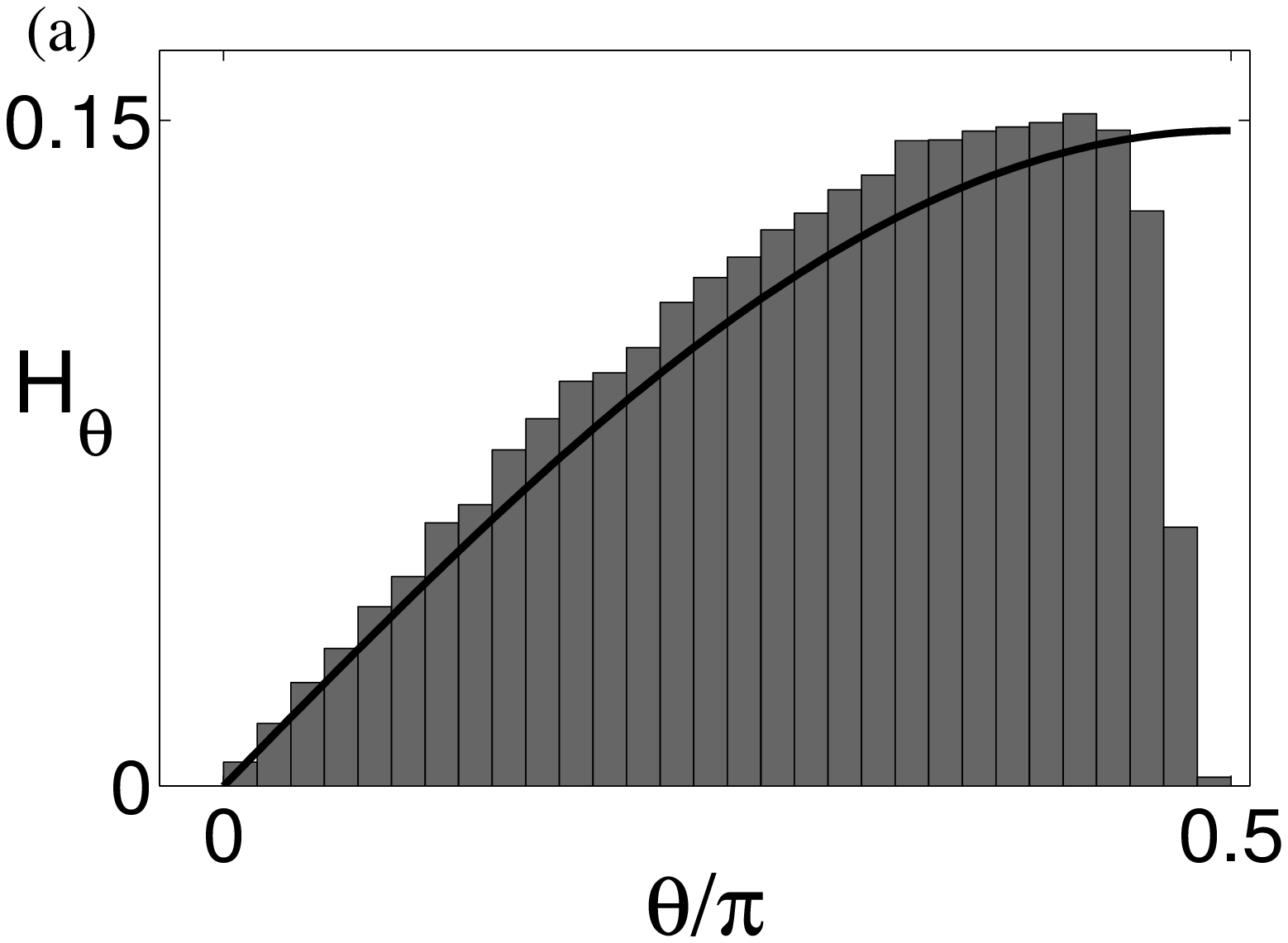, width=0.4\textwidth}
\epsfig{file=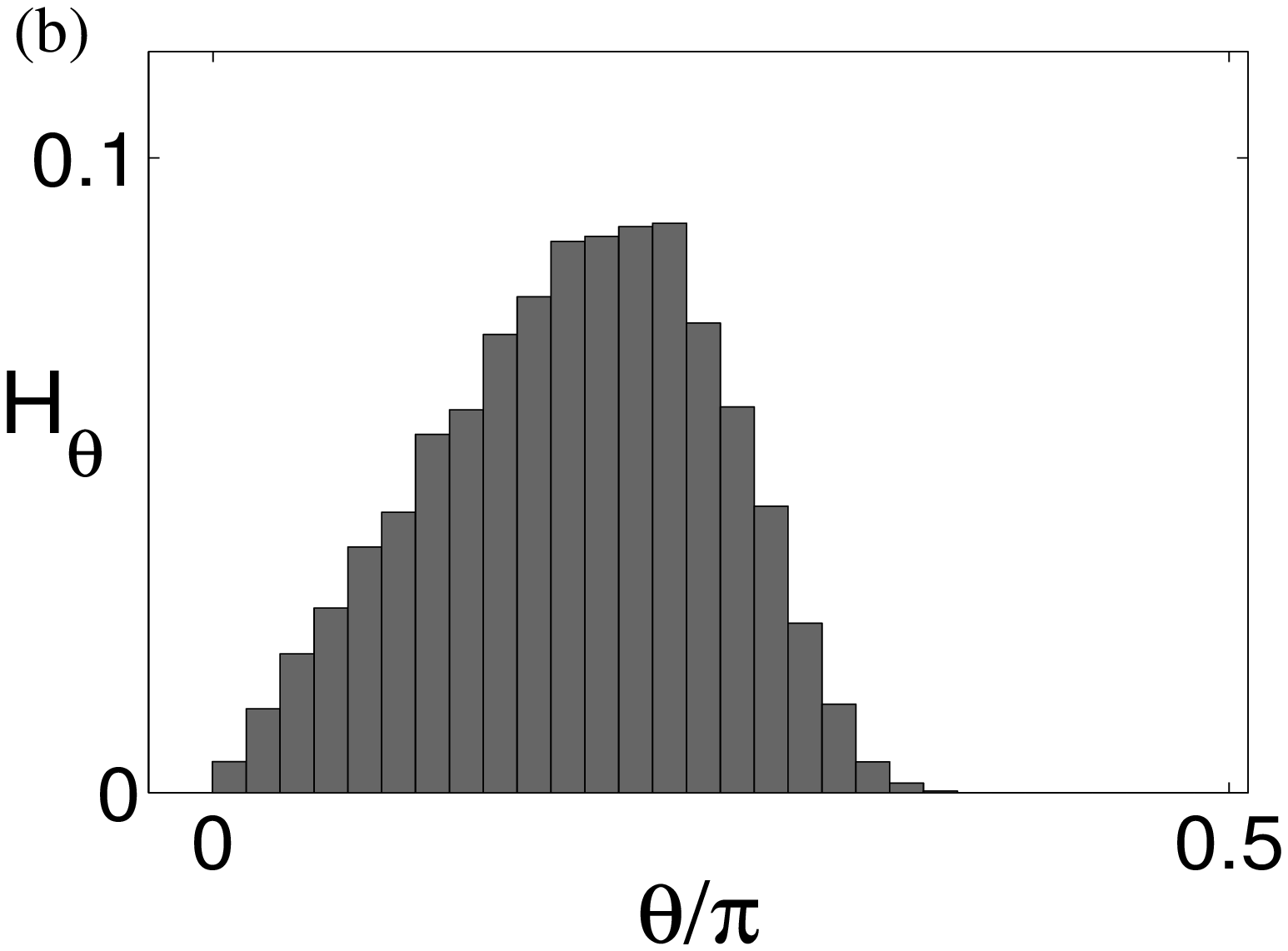, width=0.4\textwidth} 
\caption{Histogram of the orientation (relative to $\vec{E}_{ext}$) 
of the induced dipole moment for all
  resonances (a), and only the biggest ones (b). The solid line in (a) 
is obtained if the orientation of the dipole
  axis is entirely random.
  \label{fig:histo_orient} }
\end{figure}

In a plasmonic waveguide, consisting of a chain of metallic nanoparticles, a
random angle between the induced dipole moment $\vec{p}$ and the incident field
$\vec{E}_{ext}$ influences an efficient transport, because the near field of
one particle should excite the neighboring particle. Therefore designing a
plasmonic waveguide requires the knowledge of the spatial structure of the
near field.  Because a sphere has no principal axes, any arbitrarily small
perturbation will pick three axes and therefore affects the orientation of the
induced dipole. Thus, with perturbed spheres we may expect that there are only
very few cases in which $\vec{p}$ is nearly parallel to $\vec{E}_{ext}$.  To
illustrate this quantitatively, we present histograms of the angle between
$\vec{p}$ and $\vec{E}_{ext}$ in Fig.~\ref{fig:histo_orient}. In
Fig.~\ref{fig:histo_orient}(a) we consider all resonances, whereas in
Fig.~\ref{fig:histo_orient}(b) only the biggest resonance for every member is
used. The main result is that realizations with $\vec{p}$ nearly parallel to
$\vec{E}_{ext}$ are indeed negligible, and the orientation of the dipole
relatively to the external field is nearly random. This can be seen from 
the solid line in
Fig.~\ref{fig:histo_orient}(a), proportional to $\sin(\theta)$, which
corresponds to a completely random choice of the orientation of the dipoles. This
figure also illustrates that the only cases which 
are suppressed when the spheres are
perturbed are those in which $\vec{p}$ and $\vec{E}_{ext}$ are almost
perpendicular.

As expected, when only the largest resonances are considered the angle is
typically smaller because, as already pointed out, in a resonant situation only
that part of the incident field contributes that is parallel to the according
dipole axis. Moreover, the distribution of this angle, shown in
Fig.~\ref{fig:histo_orient}(b), is quite broad, which demonstrates the
substantial variability in the spatial structure of the field around our 
perturbed spheres.

Therefore, we now consider the surface modes of
perturbed {\em spheroids}. As spheroids already have at least one distinguished 
principal axis, we expect that perturbations of a spheroid do 
not have such a strong effect on the orientation of the induced dipoles. We employ
the same kind of Gaussian perturbations, but now the unperturbed particle is a
spheroid with semi-axes $R_x=1$, $R_y=1$, and $R_z=1.2$. Hence, the shape of a
particle is described by

\begin{align}
r(\theta,\phi)=&\left[(\sin \theta)^2+\left(\frac{\cos \theta}{1.2}\right)^2
  \right]^{-0.5}\\&+\sum\limits_{i=1}^n h_i
\exp\left(-0.5\left(\frac{{\rm dist}(\theta_i,\phi_i;\theta,\phi)}{w_i} \right)^2
\right).\nonumber
\end{align}

Again, $\theta_i$ and $\phi_i$ are uniformly distributed on a sphere, $h_i$ are
normally distributed with a mean value of $0.2$ and a standard deviation of
$0.1$, and the normally distributed $w_i$ have a mean value of $0.7$ and a
standard deviation of $0.3$. We consider 10000 realizations of the spheroid and
calculate the response to a constant field in $z$-direction for each. 
Here the numerical criterion $e_{1,2}\le 0.1$ leaves us with 4323 particles. 

\begin{figure}[Hhbt]
\centering
\epsfig{file=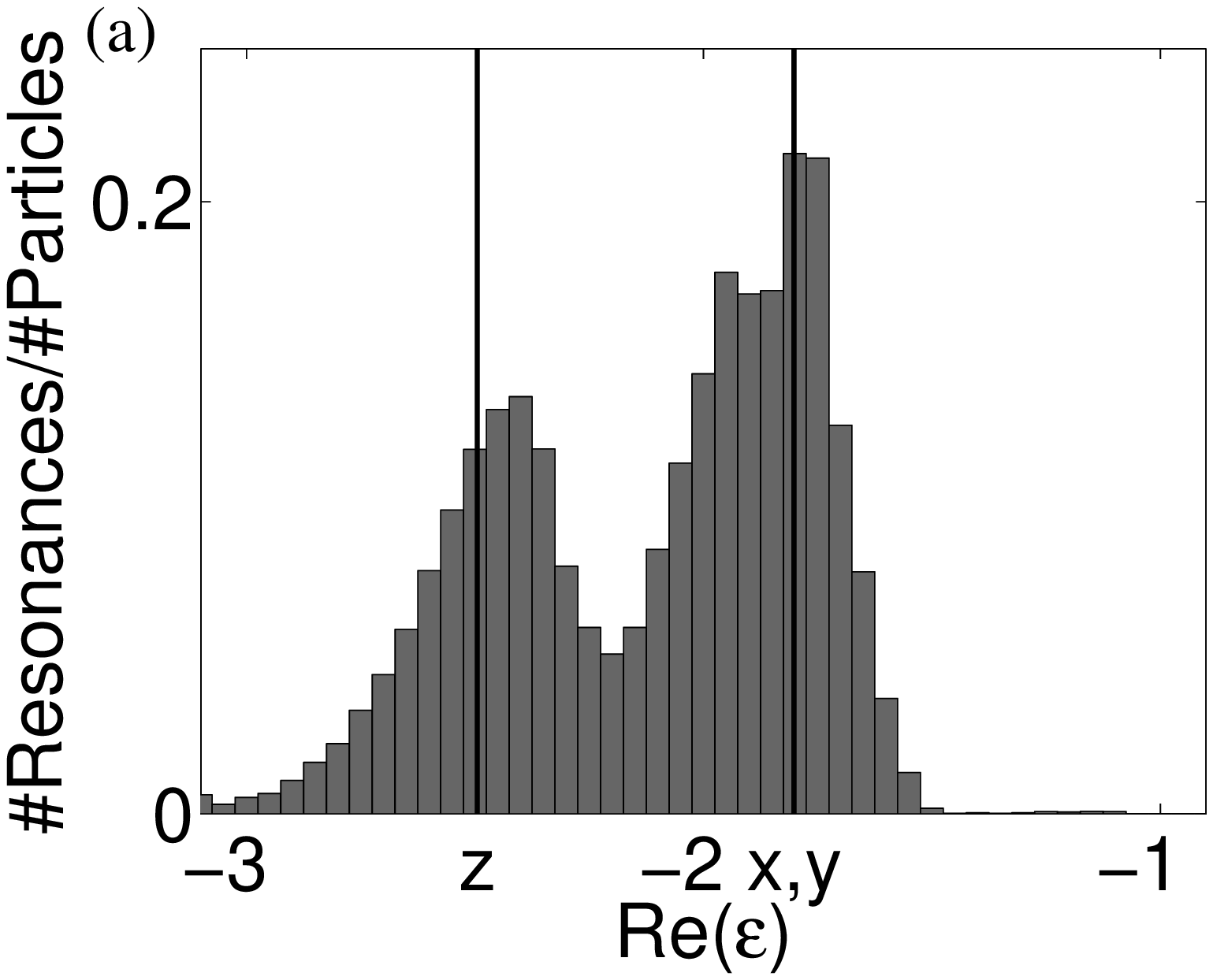, width=0.4\textwidth}
\epsfig{file=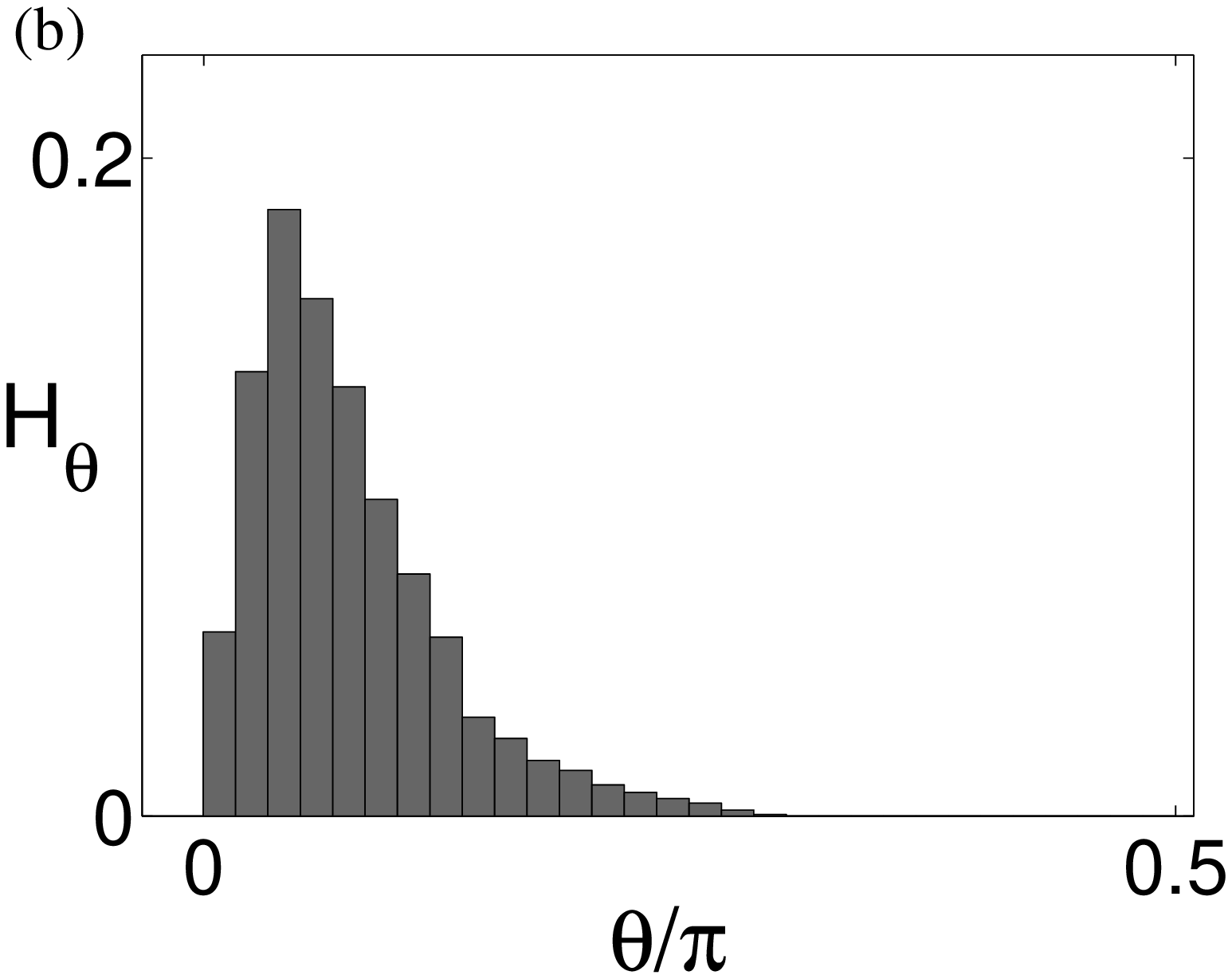, width=0.4\textwidth}
\caption{\label{fig:histo_ellip1} Location of the resonances ((a):
  all) and orientation of the induced dipole moments ((b): largest
  only) for the perturbed spheroids. The vertical lines in (a),
  labeled by {\it z} and {\it x,y}, indicates the resonant values for the
  unperturbed spheroid and an incident field along respective coordinate axes.}
\end{figure}

In Fig.~\ref{fig:histo_ellip1} we present histograms for the location
of the resonances and for the orientation of the induced
dipoles. Again the resonances are shifted to somewhat bigger values of
the permittivity. However, now the angle between the incident field
and the induced dipole moment for the biggest resonance is much
smaller, and also the distribution of the angle is narrower than in
the previous case of the perturbed spheres. This comparison can be
roughly quantified by the respective mean values, and the widths of
the distributions of the angles. We focus again on the biggest
resonances only, and determine the mean value $\overline{\theta}$ of
the angle and the interval $I_{\theta}$, centered around the mean
value, which contains two-thirds of the resonances. For the case of
the sphere this results in $\overline{\theta}=1.04$ and
$I_{\theta}=[0.65:1.43]$, whereas $\overline{\theta}=0.23$ and
$I_{\theta}=[0.09:0.37]$ for the case of spheroids.

 \section{Conclusion}

We have presented a simple-to-implement and efficient numerical scheme for
calculating important characteristics for surface modes of sphere-like
nanoparticles in the quasistatic limit, based on an expansion of the ``inner''  and
``outer'' potentials into spherical harmonics. Although the spherical harmonics do
not constitute an orthogonal basis for particles which are not exactly
spherical, and therefore encounter problems similar to those connected with the
Rayleigh hypothesis when the deviation from an exact sphere becomes too strong,
they still remain a useful system of functions, in particular so when the
boundary conditions are interpreted in an integral manner, with a 
physically motivated choice of test functions. 

We have validated this scheme against exact solutions for ellipsoids, and
against perturbation-theoretical calculations for deformed spheres.  These 
comparisons and also additional
intrinsic numerical tests both show that our method is able to yield
accurate results for the resonant permittivities and the polariziabilities even
when only quite few spherical harmonics are employed, that is, with quite small
basis sets.

This high computational efficiency allows to perform statistical
studies of large ensembles of randomly perturbed nanoparticles. This
ability is indispensable when designing, e.g., plasmonic waveguides from
nanoparticles with small, but uncontrolled fabrication-induced shape
fluctuations. On the one hand, typical effects of such imperfections on the
performance of these devices can be quantified in this manner; on the other,
admissible tolerances can be determined. While the specific distribution of
shape fluctuations employed for demonstration purposes in our example given in
Sec. IV may not be realistic, the key steps of such a large-scale statistical
analysis proceed along exactly the same route as outlined there.

\section*{Acknowledgments}
This work was supported in part by the DFG through Grant No. KI 438/8-1.
Computer power was obtained from the GOLEM I cluster of the Universit\"at
Oldenburg. We thank S.-A. Biehs, D. Grieser, M. Holthaus and O. Huth for the
fruitful discussions.

\begin{widetext}
\appendix
\section{Structure of the matrices}\label{sec:app}
The structure of the matrices in Eq.~\reff{lsys}, 
i.e.~$(M_1+\epsi M_2)U=M_3G$, is: 
\begin{equation}
{\small 
M_1{=}\left(
\begin{array}{c c c c| c c c c}
a_{0,0,0,0} & a_{1,-1,0,0} & \cdots & a_{N,N,0,0} 
&b_{0,0,0,0} & b_{1,-1,0,0} & \cdots & b_{N,N,0,0} \\
a_{0,0,1,-1} & a_{1,-1,1,-1} & \cdots &a_{N,N,1,-1} &
b_{0,0,1,-1} & b_{1,-1,1,-1} & \cdots & b_{N,N,1,-1}\\
\vdots & \vdots & \ddots & \vdots & 
\vdots &\vdots & \vdots & \ddots \\
 a_{0,0,N,N} & a_{1,-1,N,N} & \cdots &a_{N,N,N,N} 
& b_{0,0,N,N}& b_{1,-1,N,N} & \cdots &b_{N,N,N,N}\\
\hline
0 & 0 & \cdots & 0 &
d_{0,0,0,0} & d_{1,-1,0,0} & \cdots & d_{N,N,0,0} \\
\vdots & \vdots & \ddots &\vdots & 
d_{0,0,1,-1} & d_{1,-1,1,-1} & \cdots & d_{N,N,1,-1}\\
\vdots & \vdots & \ddots & \vdots & 
\vdots & \vdots & \ddots & \vdots \\
 0 & 0 & \cdots &0 & d_{0,0,N,N} & d_{1,-1,N,N} & \cdots &d_{N,N,N,N}
\end{array}
\right)},\notag
\end{equation}
\begin{equation}
M_2=\left(
\begin{array}{c | c }
\vec{0} & \vec{0}\\
\hline
(c_{l,m,l',m'}) & \vec{0} 
\end{array}
\right),\qquad  
M_3=\left(
\begin{array}{c | c }
(a_{l,m,l',m'}) & \vec{0}\\
\hline
(c_{l,m,l',m'}) & \vec{0} 
\end{array}
\right), \notag 
\end{equation}
where $\vec{0}$ is the $(N+1)^2\times(N+1)^2$ zero matrix. 
The coefficients $a_{l,m,l',m'}$, $\cdots$, $d_{l,m,l',m'}$ are defined
by projections of the boundary conditions \reff{eq:bc1},\reff{eq:bc2} 
onto the modes $r^{l'}Y^{m'}_{l'}$ on the surface of the particle
$\partial \Omega$: 
\begin{gather*}
\begin{split}
a_{l,m,l',m'}&=\int\limits_{\partial \Omega} r(s)^{l+l'}
Y^{m}_{l}(s)\overline{Y^{m'}_{l'}(s)} \, \text{d}S\,,\\
b_{l,m,l',m'}&={-}\int\limits_{\partial \Omega} r(s)^{-(l+1)+l'}
Y^{m}_{l}(s)\overline{Y^{m'}_{l'}(s)}\, \text{d}S\,,\\
c_{l,m,l',m'}&=\int\limits_{\partial \Omega} 
\partial_n \left(r(s)^{l}
Y^{m}_{l}(s)\right)r(s)^{l'}\overline{Y^{m'}_{l'}(s)} \, \text{d}S\,,\\
d_{l,m,l',m'}&=-\int\limits_{\partial \Omega} \partial_n 
\left( r(s)^{-(l+1)}
Y^{m}_{l}(s)\right)r^{l'}\overline{Y^{m'}_{l'}(s)} \, \text{d}S\,.
\end{split}
\end{gather*}
Thus, for a sphere with radius $1$ the coefficients are:
$a_{l,m,l',m'}=\delta_{ll'}\delta_{mm'}, 
b_{l,m,l',m'}=-\delta_{ll'}\delta_{mm'},$
$c_{l,m,l',m'}=l\delta_{ll'}\delta_{mm'}, 
d_{l,m,l',m'}=(l+1)\delta_{ll'}\delta_{mm'}, $
and so only four block diagonals of the matrix $M_1+\epsi
M_2$ have nonzero entries.
The vectors $U$ and $G$ in the system \reff{lsys} are defined by the
coefficients $\alpha_{l,m}$, $\beta_{l,m}$ and $\gamma_{l,m}$ of the expansion
of the potentials \reff{eq:exp_in} and \reff{eq:exp_out} as 

$$U=\left(\alpha_{0,0}, \alpha_{1,-1}, \ldots, \alpha_{N,N}, 
\beta_{0,0}, \beta_{1,-1}, \ldots, \beta_{N,N}\right)^t, \ 
G=\left(\gamma_{0,0}, \gamma_{1,-1}, \ldots, \gamma_{N,N}, 0, \ldots, 0
\right)^t.$$
\end{widetext}

\end{document}